# High-Performance Ultra-Wide-Bandgap CaSnO$_3$ Metal-Oxide-Semiconductor Field-Effect Transistors

Weideng Sun[1,†], Junghyun Koo[1,†], Donghwan Kim[2], Hongseung Lee[3], Rishi Raj[2], Chengyu Zhu[1], Kiyoung Lee[4], Andre Mkhoyan[2], Hagyoul Bae[3], Bharat Jalan[2] and Gang Qiu[1]*

[1]Department of Electrical and Computer Engineering, University of Minnesota, Twin Cities, Minneapolis, MN, 55455, USA

[2]Department of Chemical Engineering and Materials Science, University of Minnesota, Twin Cities, Minneapolis, MN, 55455, USA

[3]Department of Electronic Engineering, Jeonbuk National University, Jeonju 54896, Republic of Korea

[4]Department of Materials Science and Engineering, Hongik University, Seoul 04066, Republic of Korea

[†]These authors contributed equally to this work.

*Corresponding author. E-mail: gqiu@umn.edu

# Abstract

The increasing demand for high-voltage and high-power electronic applications has intensified the search for novel ultrawide bandgap (UWB) semiconductors. Alkaline earth stannates possess wide band gaps and exhibit the highest room-temperature electron mobilities among all perovskite oxides. Among this family, Calcium stannate ($CaSnO_3$) has the largest band gap of ~4.7 eV, holding great promise for high-power applications. However, the demonstration of $CaSnO_3$ power electronic devices is so far limited. In this work, high-performance metal-oxide-semiconductor field-effect transistor (MOSFET) devices based on La-doped $CaSnO_3$ are demonstrated for the first time. The MOSFETs exhibit an on/off ratio exceeding $10^8$, along with field-effect mobility of 8.4 $cm^2$ $V^{-1}$ $s^{-1}$ and on-state current of 30 mA $mm^{-1}$. The high performance of the $CaSnO_3$ MOSFET devices can be ascribed to the excellent metal-to-semiconductor contact resistance of 0.73 k$\Omega$·μm. The devices also show great potential for harsh environment operations, as high-temperature operations up to 400 K have been demonstrated. An off-state breakdown voltage of 1660 V is achieved, with a breakdown field of ~8.3 MV $cm^{-1}$ among the highest reported for all UWB semiconductors. This work represents significant progress toward realizing the practical application of $CaSnO_3$ in future high-voltage power electronic technologies.

## I. Introduction

The growing demand for high-voltage, high-power electronic applications has driven the search for novel wide-bandgap semiconductors. While the prevailing wide-bandgap semiconductors SiC and GaN have made significant performance advances over traditional silicon-based power electronics[1-8], their bandgap (< 4 eV) constrains further performance gains in extreme environments and ultra-high-voltage applications. Recently, hopes have been cast on UWB semiconductors, including $\beta$-$Ga_2O_3$[9-14], diamond[15,16], and AlN[17,18], which have emerged as promising candidates for the next generation of power electronic devices that offer enhanced performance characteristics such as higher breakdown strength, improved thermal stability, and superior electron transport properties. Among the promising candidates, complex oxide semiconductors, especially those with the perovskite crystal structure ($ABO_3$), have garnered considerable attention[19-21]. These perovskite oxides are a family of isostructural compounds with similar lattice parameters and a wide range of tunable electronic and optical properties, enabling versatile functionalities and applications through doping, alloying, and heterostructure engineering.

In particular, $CaSnO_3$ has emerged as a particularly compelling UWB semiconductor for power electronic device channel materials due to its attractive electronic properties and inherent stability. $CaSnO_3$ has the highest direct bandgap among the perovskite oxide family, reported to be in the range of 4.2-4.9 eV[22-24]. This wide bandgap is critical for high-voltage MOSFETs, as it is directly related to high breakdown electric field and consequently enables higher operating voltages with minimal off-state leakage current. Furthermore, $CaSnO_3$ exhibits remarkable chemical and thermal stability, maintaining its stable orthorhombic structure even under extreme temperatures and high pressures[25,26]. This robustness is a significant advantage for devices operating in harsh environments. Beyond its electrical and thermal properties, $CaSnO_3$ is highly transparent in the visible light spectrum[24,27,28], positioning it as a promising material for emerging display electronic applications[28,29].

Despite these advantageous properties, the development of high-performance $CaSnO_3$-based field-effect transistors is still facing challenges. Controllable and high-density $n$-type doping in $CaSnO_3$ was considered difficult as the shallow donors in substitution of Sn tend to self-compensate[21,28,30]. Recent advancements in developing a unique hybrid molecular beam epitaxy (hMBE) procedure have addressed these challenges, enabling reproducible $n$-type doping of $CaSnO_3$ with lanthanum (La) and achieving high electron concentrations $n_{3D}$ from $3.3 \times 10^{19}$ cm$^{-3}$ to $1.6 \times 10^{20}$ cm$^{-3}$ and the maximum room-temperature mobility of 42 cm$^2$ V$^{-1}$ s$^{-1}$ at $n_{3D} = 3.3 \times 10^{19}$ cm$^{-3}$[22]. Other thin film deposition techniques, such as pulsed laser deposition (PLD)[24,28], precipitation method[23], and chemical route[27] have also been reported to successfully synthesize $CaSnO_3$ thin films, albeit slightly inferior quality. Despite the progress in $CaSnO_3$ thin film deposition, the demonstration of power electronic devices is so far limited, with only metal-semiconductor field-effect transistor (MESFET) devices having been reported[22], where their performance was constrained by limited gate modulations. MOSFET devices, on the other hand, have not yet been achieved, due to the challenging contact and interface issues associated with UWB semiconductors. MOSFETs offer higher input impedance, lower gate leakage current, compatibility with high-voltage operations, and enhanced reliability owing to the robust oxide layer that mitigates degradation issues associated with Schottky junctions[31]. Here, we report the first demonstration and comprehensive electrical characterization of MOSFETs utilizing $CaSnO_3$ as the channel layer. Through optimized material growth and device engineering, we demonstrate top-gated enhanced-mode $CaSnO_3$ MOSFETs exhibiting large on/off ratio of over $10^8$, field-effect mobility of 8.4 $cm^2$ $V^{-1}$ $s^{-1}$, low contact resistance of 0.73 k$\Omega\cdot\mu$m, on-state current of 30 mA $mm^{-1}$. These results show great promise of $CaSnO_3$ for electronic applications in high-temperature environments. The Schottky barrier height of tens of meV is much smaller than the bandgap of $CaSnO_3$, suggesting the successful introduction of shallow donor levels by La-doping. The $CaSnO_3$ MOSFETs also show great promise for harsh environment operations, as high-temperature operations up to 400 K have been demonstrated. An off-state breakdown voltage of 1660 V is measured, corresponding to a breakdown field of ~8.3 MV $cm^{-1}$, one of the highest values among any UWB semiconductors. This achievement represents a significant step towards unlocking the full potential of $CaSnO_3$ for next-generation high-voltage and transparent electronic applications.

## II. Results and Discussion

Epitaxial La-doped $CaSnO_3$ thin film with a thickness of 15 nm was grown on (110)-oriented $GdScO_3$ substrate using hMBE. Details of the film growth are described in ref. [22] and provided in the Experimental Section. Typical doping concentrations are $3.1\times10^{19}$ ~ $1.2\times10^{20}$ $cm^{-3}$, with carrier mobility of 3.6 ~ 30.7 $cm^2$ $V^{-1}$ $s^{-1}$, as confirmed by Hall measurements (Supplementary Note 1). Reflection high-energy electron diffraction (RHEED) patterns confirm the epitaxial growth of the $CaSnO_3$ layers with smooth surfaces, as illustrated in Fig. S2. To characterize the lattice structure of the $CaSnO_3$ film, high-resolution X-ray diffraction (XRD) measurements were performed. The $2\theta$-$\omega$ scan pattern exhibits that the film peaks remain distinct from those of the substrate, as shown in Fig. 1(a). The wide-angle XRD results are presented in Fig. S3. The out-of-plane pseudocubic lattice parameter of $CaSnO_3$ was determined to be approximately 3.943 Å. This value is slightly smaller than the bulk value of ~ 3.950 Å and can be attributed to the slight in-plane tensile strain imposed by the $GdScO_3$ substrates[22,32]. The rocking curve of the film, measured near the pseudocubic (002) reflection and shown in Fig. 1(b), exhibits full widths at half maximum (FWHM) of approximately 0.084°. This value is consistent with previous reports[22], and further indicates the high structural quality of the La-doped $CaSnO_3$ thin film. The epitaxial growth of the $CaSnO_3$ film grown on a $GdSnO_3$ substrate is further directly confirmed by the high-magnification cross-sectional high-angle annular dark-field scanning transmission electron microscopy (HAADF-STEM) as shown in Fig. 1(c).

To demonstrate the potential of $CaSnO_3$ for MOSFET applications, we fabricated top-gated MOSFET devices utilizing $CaSnO_3$ as the channel material. The fabrication process flow is illustrated in Figs. 2(a) to 2(f). A 15-nm-thick La-doped $CaSnO_3$ film on $GdScO_3$ substrate (Fig.

2(a)) was first isolated into a mesa structure using the reactive ion etching (RIE) process (Fig. 2(b)). Cr/Au contacts were subsequently deposited to form Ohmic contacts (Fig. 2(c)). The channel was then recessed to within the maximum depletion width (approximately 11 nm) so that the transistor can be properly turned off (Fig. 2(d)). Noted that the contact regions under metal pads are not recessed, and the relatively higher carrier concentration under the metals ensures excellent metal-semiconductor Ohmic contacts. 10 nm $HfO_2$ gate dielectric was deposited by atomic layer deposition (ALD) (Fig. 2(e)), followed by top-gate metallization (Fig. 2(f)). Further details on device fabrication are provided in the Experimental Section. Fig. 2(g) presents the optical image of the final device structure using a top-gate overlapping with source/drain geometry, which is the default structure of the transistors in the subsequent discussions (unless stated otherwise). The non-overlapping structure with spacing between the top-gate and source/drain has also been explored, as shown in Fig. S4. However, because the device operates in the enhanced mode after recessing, the spacing between the top gate and source/drain contacts adds a large series resistance (inset of Fig. S4(b)), limiting the transistor performance. The low-magnification cross-sectional HAADF-STEM image of the MOS gate stack (Fig. 2(h)) shows atomically sharp, epitaxial, pristine interfaces between the $GdScO_3/CaSnO_3$ and $CaSnO_3/HfO_2$ layers. The corresponding energy dispersive X-Ray spectroscopy (STEM-EDX) elemental mapping reveals clear elemental boundaries, indicating uniform composition and thickness.

Fig. 3(a) shows the transfer characteristics of a typical $CaSnO_3$ MOSFET device measured at room temperature, with the $CaSnO_3$ channel thickness ($t$) of 11 nm, channel width ($W$) of 30 μm, and channel length defined as source-to-drain spacing ($L_{DS}$) of 5 μm. From the semilog plots of the drain current ($I_{DS}$) versus the gate-to-source voltage ($V_{GS}$) at different drain-to-source voltages ($V_{DS}$), the device exhibits $n$-type metal-oxide semiconductor (NMOS) behavior, with a large on/off ratio of $10^8$ observed. The subthreshold swing (SS) is extracted to be approximately 281 mV dec$^{-1}$. Dual sweep curves are shown in Fig. S5, with a small hysteresis observed, likely due to the interfacial trap states created during the channel recess process.

The linear plots of $I_{DS}$-$V_{GS}$ curves are shown as solid lines in Fig. 3(b). At low $V_{DS}$ regime, $I_{DS}$ exhibits a linear dependence on $V_{GS}$ and can be expressed as

$$I_{DS} = \mu_{FE} C_{ox} \frac{W}{L_{DS}} (V_{GS} - V_T) V_{DS}, \quad (1)$$

where $\mu_{FE}$ is the field-effect mobility of the $CaSnO_3$ layer, and $V_T$ is the threshold voltage. The corresponding transconductance $g_m$ as a function of $V_{GS}$, defined as

$$g_m = \frac{dI_{DS}}{dV_{GS}}, \quad (2)$$

are also shown in Fig. 3(b) with open circles. A peak $g_m$ of ~ 4 mS mm$^{-1}$ is measured at $V_{DS}$ = 2 V. A positive $V_T$ of 1.44 V is extracted from the inset of Fig. 3(b) at $V_{DS}$ = 0.1 V (green solid line), suggesting the MOSFET operates in the enhanced mode. The $\mu_{FE}$ = 8.4 cm$^2$ V$^{-1}$ s$^{-1}$ was extracted from the linear fitting of the $I_{DS}$-$V_{GS}$ curve with $V_{DS}$ = 0.1 V (inset of Fig. 3(b), detailed calculation in Supplementary Note 5), which is lower than the Hall mobility of 29 cm$^2$ V$^{-1}$ s$^{-1}$ (details on mobility extraction described in Supplementary Note 1). This can be understood because (1) the field-effect mobility typically underestimates carrier mobility[33]; (2) the channel recess process increases surface scattering and further degrades channel mobility.

The $I_{DS}$-$V_{DS}$ output characteristics of the CaSnO$_3$ MOSFET device are shown in Fig. 3(c). A large on-state current of about 30 mA mm$^{-1}$ is achieved, which is nearly two orders of magnitude higher than the previous reported value in MESFET devices[22], owing to better gate modulation through high-$\kappa$ dielectrics and, more importantly, lower metal-semiconductor contact resistance. The linear trend of $I_{DS}$-$V_{DS}$ curves in the low-bias regime indicates the formation of Ohmic-like contacts, which is quantitatively confirmed by a low contact resistance as will be discussed later. The on-state resistance ($R_{on}$) is extracted to be 108 kΩ·μm, which converts to a specific on-resistance $R_{on,sp}$ = 5.4 mΩ·cm$^2$. Such excellent electrical performance is highly reproducible and has been observed in many other devices, as shown in Table S1.

To further understand the carrier scattering mechanism, interface qualities, and contact resistances, we performed temperature-dependent $I_{DS}$-$V_{GS}$ measurements on another CaSnO$_3$ MOSFET device with $t$ = 11 nm, $W$ = 40 μm, and $L_{DS}$ = 5 μm. Figure 4(a) presents the $I_{DS}$-$V_{GS}$ curves at $V_{DS}$ = 1 V under different temperatures from 25 to 400 K. Dual sweep curves are shown in Fig. S6(a). The SS at different temperatures was extracted from Fig. 4(a), which approximately follows a linear temperature dependence (Fig. 4(b)), as expected from the equation

$$\mathrm{SS} = \frac{dV_{GS}}{d\log_{10}(I_{DS})} = \ln 10 \frac{k_B T}{q}(1 + \frac{C_{CSO} + C_{it}}{C_{ox}}), \tag{3}$$

where $k_B$ is the Boltzmann constant, $T$ is the absolute temperature, $q$ is the electron charge, $C_{CSO}$, $C_{it}$, and $C_{ox}$ are the capacitances of the CaSnO$_3$ layer, the interface traps, and the HfO$_2$ layer, respectively. Using the relative dielectric constants of $\varepsilon_{CSO}$ ~ 15 for CaSnO$_3$[22] and $\varepsilon_{ox}$ ~ 18 for HfO$_2$ layer with a thickness of 10 nm, the interface trap capacitance $C_{it}$ is calculated to be approximately 4.63×10$^{-6}$ F cm$^{-2}$, which translates to a interface trap density of $D_{it} = C_{it} / q^2$ ~ 2.89×10$^{13}$ (cm$^2$eV)$^{-1}$. The relatively high $D_{it}$ may be attributed to the rough CaSnO$_3$ surface after etching, and the defective self-oxidized AlO$_x$ seeding layer to facilitate ALD deposition (see Methods). From the temperature dependence of $I_{DS}$-$V_{GS}$ curves with $V_{DS}$ = 0.1 V (Fig. S6(b)), the field-effect mobility at different temperatures with $V_{DS}$ = 0.1 V can be extracted according to Supplementary Note 5, and the result is shown in Fig. 4(c). The mobility lowers with reduced temperature, and the overall trend follows the relationship of $\mu_{FE} \propto T^{1.5}$. This suggests that the impurity scattering is the dominant scattering mechanism, which is reasonable given the high doping concentration in the channel. Notably, as shown in Fig. 4a, the transistor is still functional at 400 K, despite slightly increased off-state current due to gate leakage. A prominent hysteresis behavior of $I_{DS}$-$V_{GS}$ curves is observed at temperatures above 300 K, as shown in Figs. S6(a) and S6(c), which can be attributed to thermally activated interface trap states and/or mobile ions. The increasing hysteresis at elevated temperature also obscures the extraction of $\mu_{FE}$ and SS, which explains why high temperature data in Figs. 4(b) and 4(c) deviate from the trend. These results show great promise of CaSnO$_3$ for electronic applications in high-temperature environments (Fig. S7). Further optimization on gate dielectric quality will be investigated in the future to suppress the gate leakage and improve hysteresis.

Forming Ohmic contacts is, in general, a big challenge for achieving high-performance UWB semiconductor transistors. To understand the excellent characteristics of the CaSnO$_3$ MOSFET devices, we quantitatively evaluate the metal-to-semiconductor contact resistance using the transfer length method (TLM), as shown in Fig. 5(a). Two-terminal resistances were measured with different channel lengths ranging from 3 to 20 μm. From the linear fitting, it can be deduced that the contact resistance $R_c$ of 0.73±0.02 kΩ·μm, which is among the best of UWB semiconductor contact resistance values, as presented in Table S2[1,34,35]. The Schottky barrier height $\Phi_B$ can be further quantitatively extracted from the $I_{DS}$-$V_{GS}$ at different temperatures using the equation[36-38]

$$\ln\left(\frac{I_{DS}}{T^{1.5}}\right) = -\frac{\Phi_B}{k_B T} + const., \qquad (4)$$

The Arrhenius plot of $\ln(I_{DS}/T^{1.5})$ at different $V_{GS}$ is shown in Fig. 5(b) using the data from Fig. 4(a), and the data is linearly fitted at high temperatures ($T \geq 175$ K). The $V_{GS}$-dependent effective Schottky barrier height $\Phi_B$ is then calculated from the slopes in Fig. 5(b), and the result is shown in Fig. 5(c). A low $\Phi_B$ on the order of tens of meV is extracted across the gate range from 1 V to 6 V, which is comparable to the thermal energy at 300 K, and it is consistent with the measured low contact resistance. The barrier height $\Phi_B$ is also much smaller than the bandgap of CaSnO$_3$, suggesting the successful introduction of shallow donor levels by La-doping. In fact, the linear $I_{DS}$-$V_{DS}$ at low drain bias in the output characteristics is sustained down to 25 K (same device in Fig. 3), as illustrated in Fig. 5(d). The excellent Ohmic contact at low temperatures suggests that field emission is the dominant electron transport mechanism across the heavily-doped semiconductor-metal interface rather than the temperature-dependent thermionic emission process.

Finally, to evaluate the potential for high-voltage electronic applications of CaSnO$_3$, the breakdown voltage measurement was performed in MOSFET devices with a non-overlapping geometry with a gate-to-drain length ($L_{GD}$) of 2 μm. A high breakdown voltage of 1660 V and 1200 V was measured when the gate is biased in the off state ($V_{GS}$ = -15 V) and grounded ($V_{GS}$ = 0 V), respectively, as shown in Fig. 6. The breakdown voltage at zero gate-bias is lower than that in the off state ($V_{GS}$ = -15 V) due to the presence of finite $I_{DS}$ for $V_{GS}$ = 0 V, which induces hot carrier injection to initiate breakdown through impact ionization. An exceptionally high off-state breakdown field is calculated from $V_{GS}/L_{GD}$ to be ~8.3 MV cm$^{-1}$. Noted that this value reflects the average electric field across the gate-to-drain region, and the peak electric field at the breakdown point might be even higher. The off-state breakdown field stands out among all UWB semiconductors experimentally demonstrated, as indicated in Table S3. Our measured breakdown field of the CaSnO$_3$ device also approaches the theoretical limit of ~8.5 MV cm$^{-1}$ estimated from its bandgap of 4.68 eV[22], indicating the high structural and electrical quality of the CaSnO$_3$ MOSFET devices.

## III. Conclusions

We report the first demonstration and comprehensive electrical characterization of MOSFETs utilizing hMBE-grown high-quality UWB CaSnO$_3$ as the channel layer. Through optimized material growth and device engineering, we achieved top-gated enhanced-mode CaSnO$_3$ MOSFETs exhibiting a large on/off ratio of over 10$^8$, field-effect mobility of 8.4 cm$^2$ V$^{-1}$ s$^{-1}$, and on-state current of 30 mA mm$^{-1}$. The high performance of the CaSnO$_3$ MOSFET devices can be attributed to the lowest UWB semiconductor contact resistance of 0.73 kΩ·µm. The Schottky barrier height of tens of meV is much smaller than the bandgap of CaSnO$_3$, suggesting the successful introduction of shallow donor levels by La-doping. The CaSnO$_3$ MOSFETs also show great promise for harsh environment operations, as high-temperature operations up to 400 K have been demonstrated. A breakdown field of ~8.3 MV cm$^{-1}$ is one of the highest among UWB semiconductors. This work represents a critical step in realizing the practical application of CaSnO$_3$ in future high-voltage and high-frequency power electronic technologies.

## IV. Experimental Section

**Film Growth.** The La:CaSnO$_3$ thin film was grown on a single-crystal 5 mm × 5 mm GdScO$_3$ (110) substrate (CrysTec GmbH, Germany) by using hybrid molecular beam epitaxy (hMBE). Lanthanum and calcium were supplied via conventional effusion cells, and oxygen was introduced by inductively coupled RF plasma. A vapor inlet system is employed to inject hexamethyldititn (HMDT) as a metal-organic precursor for tin source. The substrate was cleaned *in situ* for 25 minutes using 250 watts RF oxygen plasma at an oxygen background pressure of 5 × 10$^{-6}$ Torr and the substrate temperature of 700 °C (measured by a floating thermocouple) prior to film deposition. The film was subsequently grown under the same oxygen plasma conditions and substrate temperature as used in oxygen cleaning of the substrate. During growth, calcium was evaporated at 443 °C to achieve a beam equivalent pressure (BEP) of 5.0 × 10$^{-9}$ Torr, and the Lanthanum effusion cell temperature was fixed at 1100 °C to achieve the desired carrier concentrations in the film. Tin was supplied by HMDT, which offers both high volatility and high oxygen reactivity to help achieve cation and oxygen stoichiometry, respectively. HMDT was set to be supplied at a BEP of 4.2 × 10$^{-6}$ Torr. These conditions achieved a growth rate of 11 nm/h. The reader is referred to the previous publication for a detailed description of La:CaSnO$_3$ film growth[22].

**Film Characterization.** The X-ray diffraction (XRD), rocking curves, and X-ray reflectometry (XRR) experiments were performed on a Rigaku Smartlab XE high-resolution diffractometer with Cu $K_{α1}$ radiation (wavelength 1.5406 Å). Film thicknesses were fitted and extracted from the XRR results.

Samples for STEM study were prepared using a focused-ion beam (FIB) performed on a FEI Helios NanoLab G4 dual-beam system. The FIB was operated with a Ga-ion beam at 30 kV and an electron beam at 5 kV. An amorphous carbon (am-C) layer of ~50 nm was deposited on the sample as a protective layer before FIB milling. An additional 2 μm protective layer of am-C was deposited on the region of interest. The specimens were investigated with a FEI Titan G2 60-300 (S)TEM microscope equipped with a CEOS DCOR probe corrector, a monochromator, and a Super-X EDX spectrometer. The microscope was operated at 200 kV with a STEM incident probe convergence angle of 25.5 mrad and a probe current of 80 pA. HAADF-STEM images were acquired at a detector inner and outer collection of angles of 55 and 200 mrad, respectively. The specimens were further investigated for elemental mapping with a ThermoFisher Scientific Talos F200X STEM equipped with a Super-X EDX spectrometer. The microscope was operated at 200 kV with a probe convergence angle of 10.4 mrad and a current of 250 pA. HAADF-STEM images were acquired with detector inner and outer collection of angles of 46 and 200 mrad, respectively. The STEM-EDX elemental maps were analyzed with ThermoFisher Scientific Velox software.

**Device Fabrication.** The metal-oxide-semiconductor field-effect transistor (MOSFET) devices were fabricated beginning with channel isolation using optical lithography (Karl Suss MA6). Subsequently, reactive ion etching (RIE) with $BCl_3$ using an inductively coupled plasma (ICP) system was performed for 6 minutes at an etch rate of approximately 10 nm min$^{-1}$ to form the mesa structures. In the second lithography step, the source and drain electrodes were patterned to contact the $CaSnO_3$ channels, and Cr (50 nm)/Au (50 nm) electrodes were deposited by electron-beam evaporation. A channel recess etching was then performed using the same etching recipe as channel isolation to reduce the channel thickness below the maximum depletion width. Afterwards, 1.5 nm of aluminum was evaporated and naturally oxidized as a seeding layer for gate insulators to avoid short circuits between the gate and the source/drain. An insulating $HfO_2$ gate dielectric layer with a thickness of approximately 10 nm was deposited by ALD at 150 °C. The third lithography step was then performed to define the gate electrodes on the channel, followed by the deposition of Cr (50 nm)/Au (50 nm) gate electrodes using electron-beam evaporation and a subsequent solvent lift-off process.

**Electrical Measurements.** Room-temperature electrical characterization of the MOSFET devices was performed using a probe station equipped with two Keithley 2450 sourcemeter units. Temperature-dependent electrical transport measurements and Hall mobility measurements were conducted in a Quantum Design Dynacool cryostat system over a temperature range of 400 to 1.7 K and a magnetic field range of ±9 T. Hall measurements were performed on as-grown films with van der Pauw geometry through direct wire bonds onto the film. The breakdown characteristics were measured using a Keysight B1506A high-voltage parameter analyzer. During the breakdown voltage measurement, the source was grounded and the gate was biased, while the drain voltage was swept from 0 V to a high voltage to extract the gate-to-drain breakdown field.

**Figures**

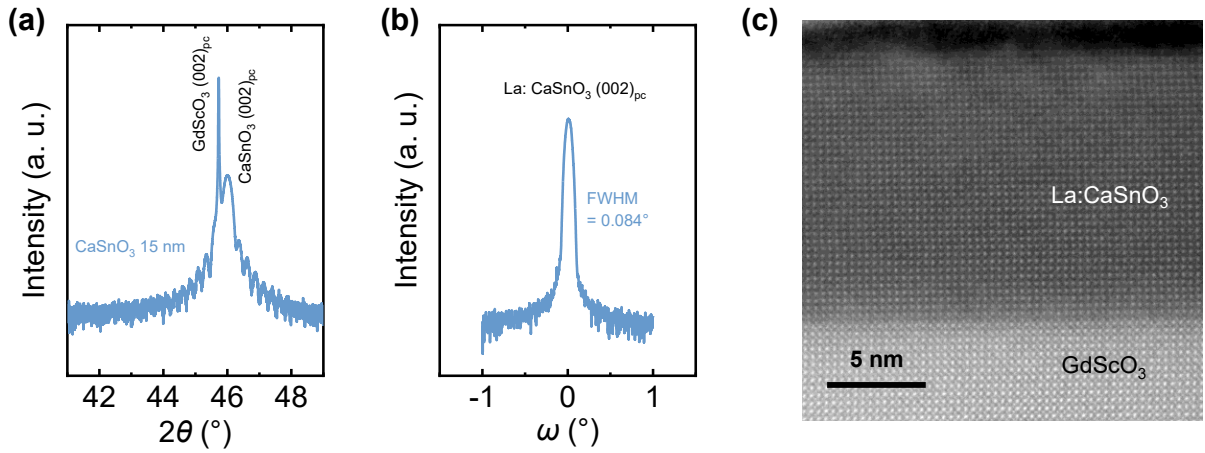

**Figure 1.** Fundamental characteristics of the La-doped CaSnO$_3$ film grown on GdScO$_3$ (110) substrate. (a) The high-resolution XRD pattern with 2$\theta$-$\omega$ scan of the sample. (b) The rocking curve of the sample measured at the pseudocubic (002) reflections. FWHM is illustrated near the curve. (c) A high-magnification HAADF-STEM image showing epitaxial growth on the clean interface between GdScO$_3$/CaSnO$_3$ layers.

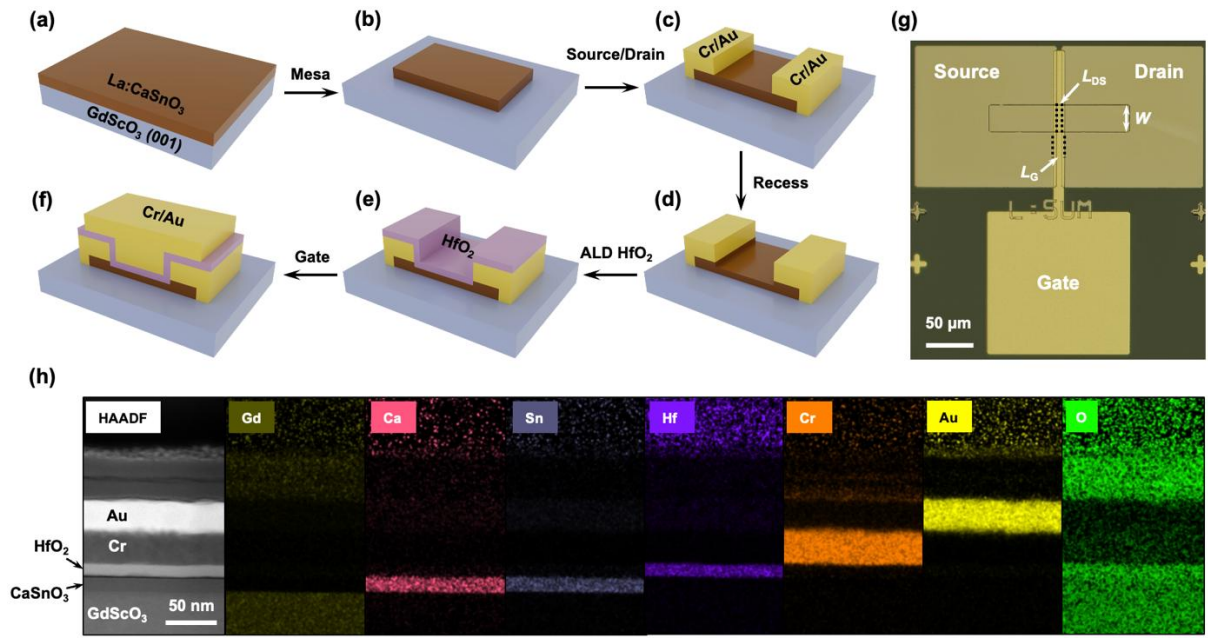

**Figure 2.** Fabrication of the La-doped CaSnO$_3$ MOSFET. (a)-(f) Fabrication process flow of the MOSFET devices. (g) An optical image of the fabricated overlapping device. The channel width ($W$), source-to-drain spacing ($L_{DS}$), and gate length ($L_G$) are illustrated. (h) A low-magnification HAADF-STEM image with EDX elemental maps. The scale bar is 50 nm.

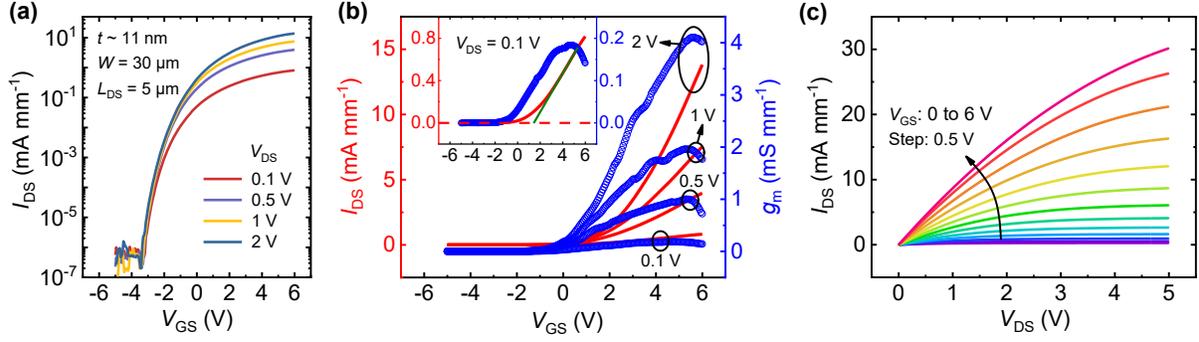

**Figure 3.** The electrical characterizations of a CaSnO$_3$ MOSFET device with $W$ = 30 μm, $L_{DS}$ = 5 μm, and $t \sim$ 11 nm. (a) Semilog plots of drain current $I_{DS}$ versus gate-to-source voltage $V_{GS}$ with different drain-to-source voltages. (b) Linear plots of $I_{DS}$-$V_{GS}$ characteristics (red curves) and extracted transconductance $g_m$-$V_{GS}$ (open circles). The inset shows the enlarged results with $V_{DS}$ = 0.1 V. (c) The output curve $I_{DS}$ versus $V_{DS}$ of the same device in (a) and (b) for different $V_{GS}$ from 0 to 6 V with a step of 0.5 V.

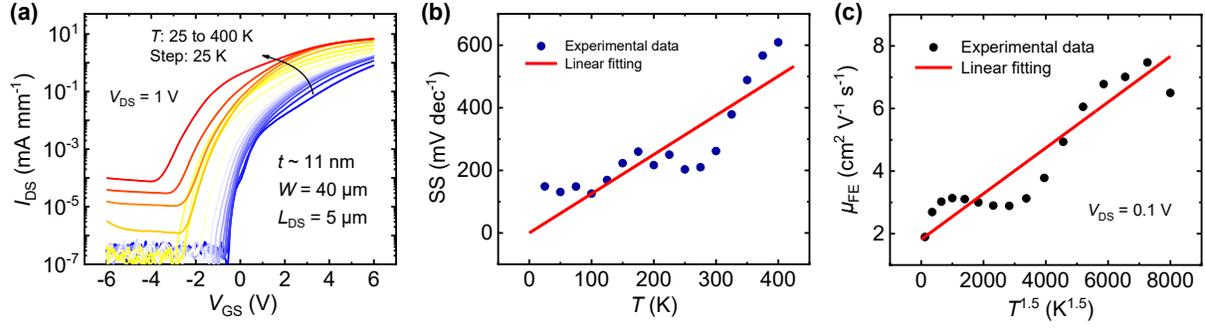

**Figure 4.** Temperature-dependent $I_{DS}$-$V_{GS}$ results with $W = 40$ μm, $L_{DS} = 5$ μm, and $t \sim 11$ nm. (a) Semilog plots of the $I_{DS}$-$V_{GS}$ curves with $V_{DS} = 1$ V and $V_{GS}$ sweeping from -6 to 6 V under different temperatures from 25 to 400 K with a step of 25 K. (b) Subthreshold swings (SS) versus temperature extracted according to Eq. (2). The blue circles and red line correspond to the experimental data and the linear fitting curve, respectively. (c) Temperature-dependent field-effect mobility $\mu_{FE}$ plotted versus $T^{1.5}$. The black circles and red line correspond to the experimental data and the linear fitting curve, respectively.

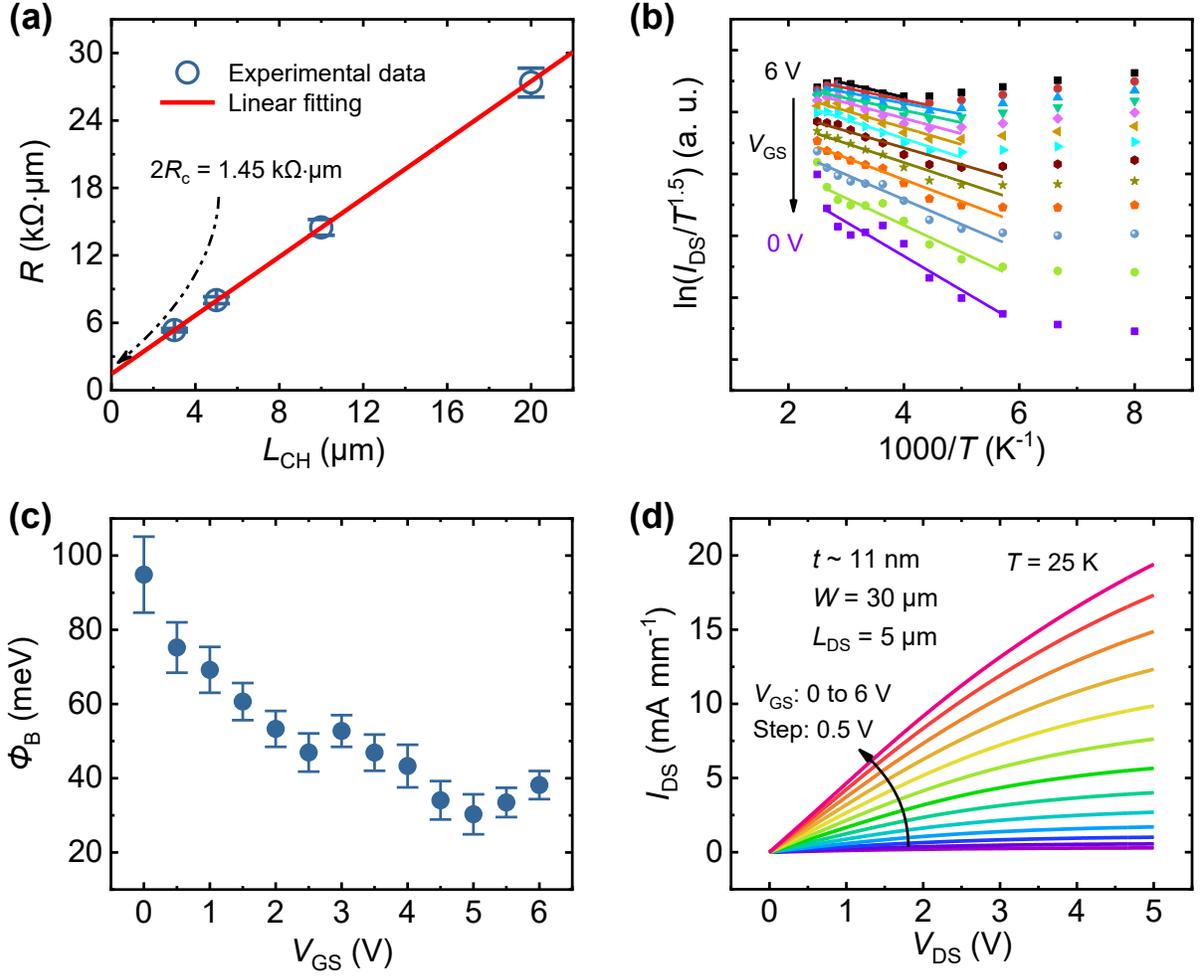

**Figure 5.** Metal-to-semiconductor contact resistance. (a) The extraction of contact resistance using the transfer length method (TLM) with channel lengths ranging from 3 to 20 μm. Each data point represents the statistical average over 10 devices, and the error bar indicates the standard deviation of the resistance. The blue circles and red line correspond to the experimental data and the linear fitting results. The intercept of the linear fitting is 1.45 kΩ·μm. (b) The Arrhenius plot at different gate biases extracted from Fig. 4(a). The data is linearly fitted at high temperatures ($T \geq 175$ K). (c) The extracted Schottky barrier height at various $V_{GS}$. (d) $I_{DS}$ versus $V_{DS}$ with $W = 30$ μm, $L_{DS} = 5$ μm, and $t \sim 11$ nm (same device in Fig. 3) for different $V_{GS}$ from 0 to 6 V with a step of 0.5 V at 25 K.

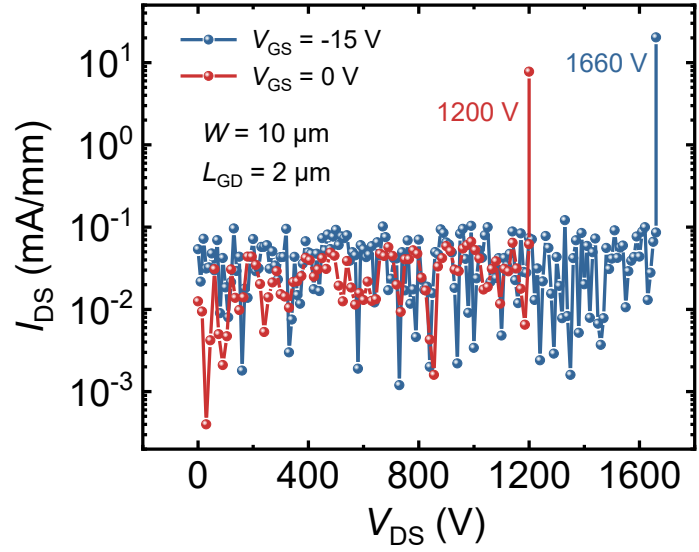

**Figure 6.** Breakdown voltage characteristics of CaSnO$_3$ MOSFETs at $V_{GS}$ = -15 V (blue) and $V_{GS}$ = 0 V (red), respectively, in a device with $W$ = 10 μm and $L_{GD}$ = 2 μm.

**Supporting Information**

Supporting Information is available from the Wiley Online Library or from the author.

**Acknowledges**

W.S. and J.K. contributed equally to this work. G. Qiu acknowledges the Department of Electrical and Computer Engineering and the College of Science and Engineering at the University of Minnesota, Twin Cities for the start-up funding support. This research was supported by the Nano & Material Technology Development Program through the National Research Foundation of Korea (NRF) funded by Ministry of Science and ICT (RS-2024-00451699). MBE growth (D.K. and B.J.) was supported by the National Science Foundation (NSF) through award number DMR-2306273. Film growth was performed using instrumentation funded by AFOSR DURIP awards FA9550-18-1-0294 and FA9550-23-1-0085. Parts of this work were carried out at the Characterization Facility, University of Minnesota, which receives partial support from the NSF through UMNMRSEC. Portions of this work were carried out at the Minnesota Nano Center, which receives support from the NSF through the National Nanotechnology Coordinated Infrastructure (NNCI) under Award No. ECCS-2025124. The electron microscopy portion of this work by R.R and K.A.M. was supported by NSF through award No. DMR-2309431, carried out in the Characterization Facility of the University of Minnesota, which receives partial support from the NSF through the MRSEC program under award DMR-2011401.

**Conflict of Interest:** The authors have no conflicts to disclose.

**Data Availability Statement**

The data that support the findings of this study are available from the corresponding author upon reasonable request.

# References

[1] Junzhe Kang, Kai Xu, Hanwool Lee, Souvik Bhattacharya, Zijing Zhao, Zhiyu Wang, R. Mohan Sankaran, and Wenjuan Zhu, Applied Physics Letters **122** (8) (2023).

[2] M. Roschke and F. Schwierz, IEEE Transactions on Electron Devices **48** (7), 1442 (2001).

[3] H. Li, S. Zhao, X. Wang, L. Ding, and H. A. Mantooth, IEEE Transactions on Power Electronics **38** (8), 9731 (2023).

[4] Zifeng Ni, Zongyu Chen, Guomei Chen, Xueyu Lu, Guohua Chen, and Ming Liu, Applied Physics A **131** (3), 206 (2025).

[5] E. Akso, H. Collins, C. Clymore, W. Li, M. Guidry, B. Romanczyk, C. Wurm, W. Liu, N. Hatui, R. Hamwey, P. Shrestha, S. Keller, and U. K. Mishra, IEEE Microwave and Wireless Technology Letters **33** (6), 683 (2023).

[6] Lili Han, Xiansheng Tang, Zhaowei Wang, Weihua Gong, Ruizhan Zhai, Zhongqing Jia, and Wei Zhang, Crystals **13** (6), 911 (2023).

[7] G. Lyu, J. Sun, J. Wei, and K. J. Chen, IEEE Transactions on Power Electronics **38** (10), 12648 (2023).

[8] Maciej Matys, Kazuki Kitagawa, Tetsuo Narita, Tsutomu Uesugi, Jun Suda, and Tetsu Kachi, Applied Physics Letters **121** (20) (2022).

[9] S. J. Pearton, Jiancheng Yang, Patrick H. Cary, IV, F. Ren, Jihyun Kim, Marko J. Tadjer, and Michael A. Mastro, Applied Physics Reviews **5** (1) (2018).

[10] Chenlu Wang, Jincheng Zhang, Shengrui Xu, Chunfu Zhang, Qian Feng, Yachao Zhang, Jing Ning, Shenglei Zhao, Hong Zhou, and Yue Hao, Journal of Physics D: Applied Physics **54** (24), 243001 (2021).

[11] Hardhyan Sheoran, Vikram Kumar, and Rajendra Singh, ACS Applied Electronic Materials **4** (6), 2589 (2022).

[12] Youngseo Park, Jiyeon Ma, Geonwook Yoo, and Junseok Heo, Nanomaterials **11** (2), 494 (2021).

[13] H. Zhou, M. Si, S. Alghamdi, G. Qiu, L. Yang, and P. D. Ye, IEEE Electron Device Letters **38** (1), 103 (2017).

[14] Lin-Qing Zhang, Wan-Qing Miao, Xiao-Li Wu, Jing-Yi Ding, Shao-Yong Qin, Jia-Jia Liu, Ya-Ting Tian, Zhi-Yan Wu, Yan Zhang, Qian Xing, and Peng-Fei Wang, Inorganics **11** (10), 397 (2023).

[15] Hiromitsu Kato, Hitoshi Umezawa, Norio Tokuda, Daisuke Takeuchi, Hideyo Okushi, and Satoshi Yamasaki, Applied Physics Letters **93** (20) (2008).

[16] Wenchao Zhang, Benjian Liu, Sen Zhang, Xiaohui Zhang, Pengfei Qiao, Bo Liang, Saifei Fan, Tao Su, Kang Liu, Bing Dai, and Jiaqi Zhu, The Journal of Physical Chemistry Letters **15** (36), 9301 (2024).

[17] Dhanu Chettri, Ganesh Mainali, Haicheng Cao, Juan Huerta Salcedo, Mingtao Nong, Mritunjay Kumar, Saravanan Yuvaraja, Xiao Tang, CheHao Liao, and Xiaohang Li, Journal of Physics D: Applied Physics **58** (3), 035104 (2025).

[18] H. Fu, I. Baranowski, X. Huang, H. Chen, Z. Lu, J. Montes, X. Zhang, and Y. Zhao, IEEE Electron Device Letters **38** (9), 1286 (2017).

[19] H. Y. Hwang, Y. Iwasa, M. Kawasaki, B. Keimer, N. Nagaosa, and Y. Tokura, Nature Materials **11** (2), 103 (2012).

[20] Abhinav Prakash and Bharat Jalan, Advanced Materials Interfaces **6** (15), 1900479 (2019).

[21] William Nunn, Tristan K. Truttmann, and Bharat Jalan, Journal of Materials Research **36** (23), 4846 (2021).

[22] Fengdeng Liu, Prafful Golani, Tristan K. Truttmann, Igor Evangelista, Michelle A. Smeaton, David Bugallo, Jiaxuan Wen, Anusha Kamath Manjeshwar, Steven J. May, Lena F. Kourkoutis, Anderson Janotti, Steven J. Koester, and Bharat Jalan, ACS Nano **17** (17), 16912 (2023).

[23] Manh Hoang Tran, Taehyun Park, and Jaehyun Hur, ACS Applied Materials & Interfaces **13** (11), 13372 (2021).

[24] Qinzhuang Liu, Feng Jin, Bing Li, and Lei Geng, Journal of Alloys and Compounds **717**, 55 (2017).

[25] S. A. T. Redfern, C. J. Chen, J. Kung, O. Chaix-Pluchery, J. Kreisel, and E. K. H. Salje, Journal of Physics: Condensed Matter **23** (42), 425401 (2011).

[26] Benjamin W. Schneider, Liu Wei, and Baosheng and Li, High Pressure Research **28** (3), 397 (2008).

[27] Hamza Shaili, Elmehdi Salmani, Mustapha Beraich, Mustapha Zidane, M'hamed Taibi, Mustapha Rouchdi, Hamid Ez-Zahraouy, Najem Hassanain, and Ahmed Mzerd, ACS Omega **6** (48), 32537 (2021).

[28] Mian Wei, Hai Jun Cho, and Hiromichi Ohta, ACS Applied Electronic Materials **2** (12), 3971 (2020).

[29] Fengdeng Liu, Zhifei Yang, David Abramovitch, Silu Guo, K. Andre Mkhoyan, Marco Bernardi, and Bharat Jalan, Science Advances **10** (44), eadq7892 (2024).

[30]   L. Weston, L. Bjaalie, K. Krishnaswamy, and C. G. Van de Walle, Physical Review B **97** (5), 054112 (2018).

[31]   S. M. Sze and Kwok K. Ng, *Physics of Semiconductor Devices*, 3 ed. (John Wiley & Sons, Inc., Hoboken, New Jersey, 2007).

[32]   H. Mountstevens Elizabeth, J. Paul Attfield, and A. T. Redfern Simon, Journal of Physics: Condensed Matter **15** (49), 8315 (2003).

[33]   Yuan Taur and Tak H. Ning, *Fundamentals of Modern VLSI Devices*, 3 ed. (Cambridge University Press, Cambridge, 2021).

[34]   Joo Hee Jeong, Seung Wan Seo, Dongseon Kim, Seong Hun Yoon, Seung Hee Lee, Bong Jin Kuh, Taikyu Kim, and Jae Kyeong Jeong, Scientific Reports **14** (1), 10953 (2024).

[35]   Y. Lv, H. Liu, X. Zhou, Y. Wang, X. Song, Y. Cai, Q. Yan, C. Wang, S. Liang, J. Zhang, Z. Feng, H. Zhou, S. Cai, and Y. Hao, IEEE Electron Device Letters **41** (4), 537 (2020).

[36]   Jingli Wang, Qian Yao, Chun-Wei Huang, Xuming Zou, Lei Liao, Shanshan Chen, Zhiyong Fan, Kai Zhang, Wei Wu, Xiangheng Xiao, Changzhong Jiang, and Wen-Wei Wu, Advanced Materials **28** (37), 8302 (2016).

[37]   Jen-Ru Chen, Patrick M. Odenthal, Adrian G. Swartz, George Charles Floyd, Hua Wen, Kelly Yunqiu Luo, and Roland K. Kawakami, Nano Letters **13** (7), 3106 (2013).

[38]   Xu Cui, En-Min Shih, Luis A. Jauregui, Sang Hoon Chae, Young Duck Kim, Baichang Li, Dongjea Seo, Kateryna Pistunova, Jun Yin, Ji-Hoon Park, Heon-Jin Choi, Young Hee Lee, Kenji Watanabe, Takashi Taniguchi, Philip Kim, Cory R. Dean, and James C. Hone, Nano Letters **17** (8), 4781 (2017).